\documentclass[aps,prc,twocolumn,showpacs,nofootinbib]{revtex4-1}
\usepackage[utf8]{inputenc}
\usepackage{graphicx}   %       for graphics
\usepackage{latexsym}   %       for special symbols
\usepackage{enumerate}
\usepackage{rotating,booktabs,multirow}
\usepackage{amsmath}
\usepackage{amsfonts}
\usepackage{amssymb}
\usepackage{bm}

\begin{document}
%============================================================================
\title{A first estimate of $\eta/s$ in Au+Au reactions at E$_{\rm lab}=1.23$~$A$GeV}
\author{Tom Reichert$^{1}$, Gabriele Inghirami$^{2,5,6}$, Marcus~Bleicher$^{1,2,3,4}$}

\affiliation{$^1$ Institut f\"ur Theoretische Physik, Goethe Universit\"at Frankfurt, Max-von-Laue-Strasse 1, D-60438 Frankfurt am Main, Germany}
\affiliation{$^2$ GSI Helmholtzzentrum f\"ur Schwerionenforschung GmbH, Planckstr. 1, 64291 Darmstadt , Germany}
\affiliation{$^3$ John von Neumann-Institut f\"ur Computing, Forschungzentrum J\"ulich,
52425 J\"ulich, Germany}
\affiliation{$^4$ Helmholtz Research Academy Hesse for FAIR, GSI Helmholtz Center, Campus Frankfurt, Max-von-Laue-Str. 12, 60438 Frankfurt, Germany}
\affiliation{$^5$ University of Jyv\"askyl\"a, Department of Physics, P.O. Box 35, FI-40014 University of Jyv\"askyl\"a, Finland}
\affiliation{$^6$ Helsinki Institute of Physics, P.O. Box 64, FI-00014 University of Helsinki, Finland}

\begin{abstract}
The HADES experiment at GSI has recently provided data on the flow coefficients $v_1,...,v_4$ for protons in Au+Au reactions at $E_{\rm lab} = 1.23$~$A$GeV (or $\sqrt{s_\mathrm{NN}}=2.4$~GeV). This data allows to estimate the shear viscosity over entropy ratio, $\eta/s$ at low energies via a coarse graining analysis of the UrQMD transport simulations of the flow harmonics in comparison to the experimental data. By this we can provide for the first time an estimate of $\eta/s\approx0.65\pm0.15$ (or $(8\pm2)\,(4\pi)^{-1}$) at such low energies.
\end{abstract}

\maketitle
\section{Introduction}
Relativistic heavy ion collisions provide an excellent laboratory to explore the properties of strongly interacting matter in great detail. While at ultra-relativistic collision energies, e.g. at RHIC and LHC, Quantum Chromodynamics (QCD) predicts the creation of a deconfined state of matter, called the Quark Gluon Plasma (QGP), one expects to explore a super dense baryonic system at lower collision energies (RHIC-BES, FAIR/NICA, or GSI). One of the key results of the exploration of QCD matter in collider experiments has been the precise determination of the QCD transport coefficients. In particular, the shear viscosity $\eta$ which is usually scaled by the entropy density $s$ of the matter is in the center of interest. At ultra-relativistic energies, the shear viscosity $\eta$ can be rather directly extracted by analyzing the Fourier expansion of the radial flow distribution. Especially, the second Fourier coefficient $v_2$ can be measured at midrapidity (e.g. at RHIC \cite{Ackermann:2000tr,Adler:2003cb}) with great precision and allows to extract $\eta/s$ from the data by comparison to hydrodynamic simulations \cite{Huovinen:2001cy,Song:2007fn,Romatschke:2007mq,Luzum:2008cw}. Typical values of $\eta/s$ at collider energies are $\eta/s=(2-4)\, (4\pi)^{-1}$. These values are near the conformal Kovtun-Son-Starinets (KSS) bound of $\eta/s=(4\pi)^{-1}$ \cite{Policastro:2001yc,Kovtun:2004de} and indicate that the QGP is among the most perfect liquids ever created. Although $\eta/s$ cannot be smaller than the KSS bound in equilibrium, recent AdS/CFT calculations point out that out-of-equilibrium values of $\eta/s$ can subceed the boundary \cite{Wondrak:2020tzt,Wondrak:2020kml}.

Unfortunately, the methods used at collider energies cannot easily be applied at very low energies. The reason is the following: At very high energies, the initial conditions for the hydrodynamic evolution can be straightforwardly defined, e.g. at midrapidity by a time independent spatial profile with a fixed eccentricity $\epsilon$ in space. The shape of the spatial profile can be extracted from Monte-Carlo Glauber calculations or from a saturation model. This initial state then expands into vacuum and creates a finally observable $v_2$ in momentum space out of the initial eccentricity. The response of the momentum space to the spatial eccentricity is governed by the viscosity. I.e. typically one has $v_n\sim\varepsilon_n\exp[-\eta/s\cdot n^2]$ \cite{Lacey:2011ug,Shuryak:2013ke}. Thus comparing $\epsilon_2$ with $v_2$ allows to extract the viscosity.

At lower energies, such an approach is not possible due to the intricate time dependence of the geometry of the colliding and expanding matter:
\begin{itemize}
\item
First of all, the expected viscous correction might be too high to allow for a meaningful hydrodynamic simulation. Estimates of the shear viscosity over entropy ratio of a moderately hot hadron gas \cite{Demir:2008tr} suggest $\eta/s=(10-100)\, (4\pi)^{-1}$ in the low energy regime.
\item
Secondly, a simple initial state (i.e. fixed in time and a with a geometry inferable from the initial nuclei) for the hydrodynamic expansion cannot be defined. The reasons for this problem are: I) The compression stage lasts very long ($\approx3-10$~fm), which means that the initial baryon currents do not quickly decouple as they do at very high energies, hindering the definition of a time independent initial state with a fixed eccentricity $\varepsilon_n$. II) Due to the slow movement of the spectators, the expanding matter sees a time dependent boundary for its expansion, the well known spectator shadowing. This leads to an emission time dependent in-plane and out-of-plane component in $v_2$. Both effects prohibit the construction of a simple hydrodynamic initial state at low energies.
\end{itemize}

To overcome these problems one can either employ a viscous multi-fluid approach \cite{Batyuk:2016qmb} to include the space and time dependent source terms for the midrapidity fluid, or one can rely on transport simulations to extract the shear viscosity to entropy density ratio. 

In the present paper we take the second route, because viscous multi-fluid hydrodynamics is not yet available. In \cite{Hillmann:2018nmd,Hillmann:2019cfr,Hillmann:2019wlt}, the UrQMD model has recently demonstrated its ability to describe the directed and elliptic flow data from HADES \cite{Kardan:2017knj,Adamczewski-Musch:2020iio} and to predict triangular and quadrangular flow which has been recently measured \cite{Adamczewski-Musch:2020iio}. This consistency means that the implemented cross sections and interactions can be used to calculate a reliable viscosity value that is in line with the observed elliptic and higher flow harmonics without the need to rely on the initial eccentricity. This allows for the first time, to extract the $\eta/s$ ratio in Au+Au collisions at E$_\mathrm{lab}=1.23$~$A$GeV directly from a dynamical hadron cascade simulation. The idea is supplemented by the investigation of the $\eta/s$ ratio in the transverse plane at different times to illustrate the relevant viscosities at the different expansion stages.

\section{Method}
For the present analysis, the Ultra-relativistic Quantum-Molecular-Dynamics (UrQMD) model \cite{Bass:1998ca,Bleicher:1999xi} is used in its recent version (v3.4) to investigate the properties of the hadronic matter created in Au+Au reactions at 1.23 $A$GeV beam energy. At this energy, the relevant degrees of freedom are hadrons, and their interaction is based on nuclear potentials and scattering cross sections. A detailed description of the model is provided in \cite{Bass:1998ca,Bleicher:1999xi}. UrQMD has been shown to provide a very good description of the collision dynamics and especially of the flow characteristics \cite{Hillmann:2018nmd,Hillmann:2019cfr,Hillmann:2019wlt} which is relevant for the present analysis. Moreover, hydrodynamics \cite{Spieles:2020zaa}, thermal model fits \cite{Cleymans:2005xv,Agakishiev:2015bwu} and previous coarse-grained UrQMD calculations \cite{Steinheimer:2016vzu} have suggested that the matter created in the model is near equilibrium, allowing to relate equilibrium quantities ($T$, $\eta$, $s$, ...) to the system. 

So far, the extraction of $\eta/s$ from transport models generally utilizes infinite matter simulations which are run sufficiently long to allow for local thermal equilibration and then employs the Green-Kubo formalism \cite{Green_1954,Kubo_1966}. In this setup, the shear viscosity is expressed in terms of the zero-frequency slope of spectral densities of stress tensor-stress tensor correlations. This method has been successfully applied to the UrQMD model in \cite{Demir:2008tr} and the SMASH model \cite{Petersen:2008dd,Rose:2017bjz}. In \cite{Teslyk:2019ioo,Zabrodin:2020jgy}, the UrQMD model has been recently used to infer the $\eta/s$ ratio in Au+Au collisions at beam energies from 10 to 40~$A$GeV by using the local energy and net-baryon densities as input for a statistical model (SM) setup.

The aim of this work, however, is to extract the $\eta/s$ ratio locally, i.e. space- and time-dependent, from a dynamical simulation. 
%---new
The first question to answer is whether the use of local equilibrium concepts such as temperature and viscosity is justified at such low energies. 
To test, if local equilibrium is also achieved at low energies, we use the same established strategy as in high energetic collisions: We ask if (1.) relativistic hydrodynamics can be used to describe the system and (2.) if the final yields can be described by thermal model fits. In addition (3.) we can compare the presented coarse graining approach with a thermal model analysis in different time steps over the course of the reaction. 

Let us start with (1.) and test if hydrodynamic models provide a reasonable description of nuclear reactions at low energies. To this aim we refer to \cite{Spieles:2020zaa} where hydrodynamic simulations for Au+Au reactions in the energy range from $E_\mathrm{lab}=1-8$~$A$GeV are shown to provide a reasonable description of the proton and pion flow and their multiplicities.

Next we test chemical equilibration (2.). To this aim we refer to the thermal model analysis' in \cite{Cleymans:2005xv} and the updates provided in \cite{Agakishiev:2015bwu}. The comparisons of the experimental yields at low collision energies ($E_\mathrm{lab}\approx 1$~$A$GeV) with the thermal model indicates a good fit of the thermal model to the data for the bulk hadronic multiplicities (deviation for more exotic states like the $\Xi$ are present, but irrelevant due to the negligible contribution of such states).

Finally, we test (3.) how the transport simulations compare to a time dependent thermal model analysis at low energies. For this, we refer to the results shown in \cite{Steinheimer:2016vzu}, where the authors used a coarse-graining procedure as well. They showed that at $E_\mathrm{lab} = 1.76$~$A$GeV, the coarse-grained simulation data on ($T(t),\mu_B(t)$) is very close to time dependent thermal model fits ($T(t),\mu_B(t)$) of the chemical yields during the relevant overlap stage of the reaction (i.e. 4~fm/c $<t<$ 12~fm/c).

To summarize, there is substantial evidence that the matter in the discussed energy regime and for massive collisions (in our case Au+Au at 1.23~$A$GeV) is to a very good approximation in local kinetic and chemical equilibrium. Thus, the notion of viscosity and temperature is justified.
%---
 The shear viscosity $\eta$ can then be extracted using the finding that $\eta$ can be well described by the following relation extracted from relativistic kinetic theory \cite{Danielewicz:1984kt,Torres-Rincon:2012sda} in Boltzmann limit: %$\eta(T)=5/16\cdot\sqrt{\pi mT}\sigma_{tr}^{-1}$
\begin{equation}\label{eta_kinetic}
    \eta(T)=\frac{5\sqrt{\pi}}{16}\frac{\sqrt{mT}}{\sigma_{tr}},
\end{equation}
with $T$ being the temperature, $m$ being the mass and $\sigma_{tr}$ being the transport-cross section%
\footnote{The transport cross section is defined by $\sigma_{tr}(s)=\int\mathrm{d}\Omega\sin^2(\theta)\frac{\mathrm{d}\sigma}{\mathrm{d}\Omega}(s,\theta)$ \cite{Plumari:2012ep}. The transport cross section $\sigma_{tr}$ is related to the total cross section $\sigma_{tot}$ for $s$-wave (isotropic) interactions through $\sigma_{tr}=2/3\,\sigma_{tot}$ and for $p$-wave interactions via $\sigma_{tr}=2/5\,\sigma_{tot}$.}
of the interacting particles. The term $\sqrt{m}/\sigma_{tr}$ generally depends on the chemical composition produced at each individual energy and is therefore collision energy dependent. Within the simulation, we extract $\langle\sqrt{m}/\sigma_{tot}\rangle (\sqrt{s_\mathrm{NN}})$ for each considered energy. Specifically, for central Au+Au collisions at 1.23~$A$GeV, we extract $\langle\sqrt{m}/\sigma_{tot}\rangle=0.0256$~$\sqrt{\mathrm{GeV}}$/mb (as expected for a nucleon dominated system with $\approx\sqrt{m_p}/\sigma_{pp}$ and $\sigma_{pp}=40$~mb). In principle one would also be able to obtain the transport cross section at each individual space time point. We omit this extremely computational time intense task here and assume that the s-wave and p-wave scattering provide reasonable upper and lower bounds for the conversion of the averaged total cross section to the transport cross section and treat this as a systematic error. The temperature $T$ and the entropy density $s$ are then extracted at each space-time point via the coarse-graining method discussed below. To validate our approach for the extraction of $\sqrt{m}/\sigma_{tr}$, $T$, $\rho_\mathrm{B}$ and $s$ we will later also compare to the improved (in equilibrium) shear viscosity calculation by Shi and Danielewicz \cite{Shi:2003np}.

To relate the shear viscosity to the entropy density and to confront their ratio with the complex dynamics encountered in the collision, we employ the UrQMD coarse-graining approach \cite{Inghirami:2019muf,Huovinen:2002im,Endres:2014zua,Endres:2015fna,Endres:2015egk,Inghirami:2018vqd,Reichert:2020yhx}. It consists in computing the temperature and the baryo-chemical potential from the average energy-momentum tensor and net-baryon current of the hadrons formed in a large set of heavy ion collision events with the same collision energy and centrality. The computation is done in cells of a fixed spatial grid at constant intervals of time. In the present study, the cells are four-cubes with spatial sides of length $\Delta x=\Delta y=\Delta z=1$~fm and $\Delta t=0.25$~fm length in time direction. First, we evaluate the net-baryon four current $j^{\mu}_{\mathrm{B}}(t,\mathbf{r})$ and the energy momentum tensor $T^{\mu\nu}(t,\mathbf{r})$ in the Eckart's frame definition \cite{Eckart:1940te}. Then, we perform a Lorentz transformation of the net-baryon current and of the energy momentum tensor into the Local Rest Frame (LRF) and compute the local baryon density $\rho_{\mathrm{B}}$ and the energy density $\varepsilon$ as:
\begin{equation} 
\rho_{\mathrm{B}}(t,\mathbf{r}) = j_{\mathrm{B,\,LRF}}^{0}(t,\mathbf{r}),\qquad \varepsilon(t,\mathbf{r})= T^{00}_{\mathrm{LRF}}(t,\mathbf{r}).
\end{equation} 
The final step in the coarse graining procedure consists in associating to each cell of the coarse grained grid the temperature $T(\varepsilon,\rho_{\mathrm{B}})$ and the baryo-chemical potential $\mu_{\mathrm{B}}(\varepsilon,\rho_{\mathrm{B}})$ through the interpolation of a tabulated Hadron Resonance Gas EoS\footnote{Here, the HG EoS is used because of the absence of QGP at the investigated energy. When applying this method to higher energies, the employed EoS should be adjusted as well.} \cite{Zschiesche:2002zr}. 

Being close to thermal equilibrium as shown in \cite{Steinheimer:2016vzu} and utilizing that the coarse-graining employs the grand-canonical ensemble allows us to extract the local entropy density $s(t,\mathbf{r})$ from the thermodynamic relation $\Omega=-PV$ via
\begin{equation}
    s(t,\mathbf{r})=\frac{\varepsilon(t,\mathbf{r})+P(t,\mathbf{r})-\mu_\mathrm{B}(t,\mathbf{r})\rho_\mathrm{B}(t,\mathbf{r})}{T(t,\mathbf{r})}.
\end{equation}
With the local entropy density in hand, we can now define the $\eta/s$ ratio with a beam energy specific $\langle\sqrt{m}/\sigma_{tr}\rangle$.

\section{Results}
In the following we use the relation $\eta(T)=5/16\sqrt{\pi m}/\sigma_{tr}\cdot\sqrt{T}$ as derived in \cite{Danielewicz:1984kt,Torres-Rincon:2012sda}. We apply it to the coarse grained simulation data of $T(\varepsilon(t,\mathbf{r}),\rho_\mathrm{B}(t,\mathbf{r}))$ at 1.23~$A$GeV and extract the space and time dependent $\eta/s$ ratio during the evolution of the reaction. Central events are selected via an impact parameter cut at $b\le 3.4$~fm. Cold cells ($T<10$~MeV) are excluded, because the mapping of $(\varepsilon,\rho_\mathrm{B})\leftrightarrow(T,\mu_\mathrm{B},s)$ through the HRG EoS is imprecise at very low temperatures especially with respect to the entropy. The coarse-grained data is based on an ensemble of $10^7$ events.

\subsection{Time evolution of $\eta/s$}
\begin{figure} [t!hb]
	\includegraphics[width=\columnwidth]{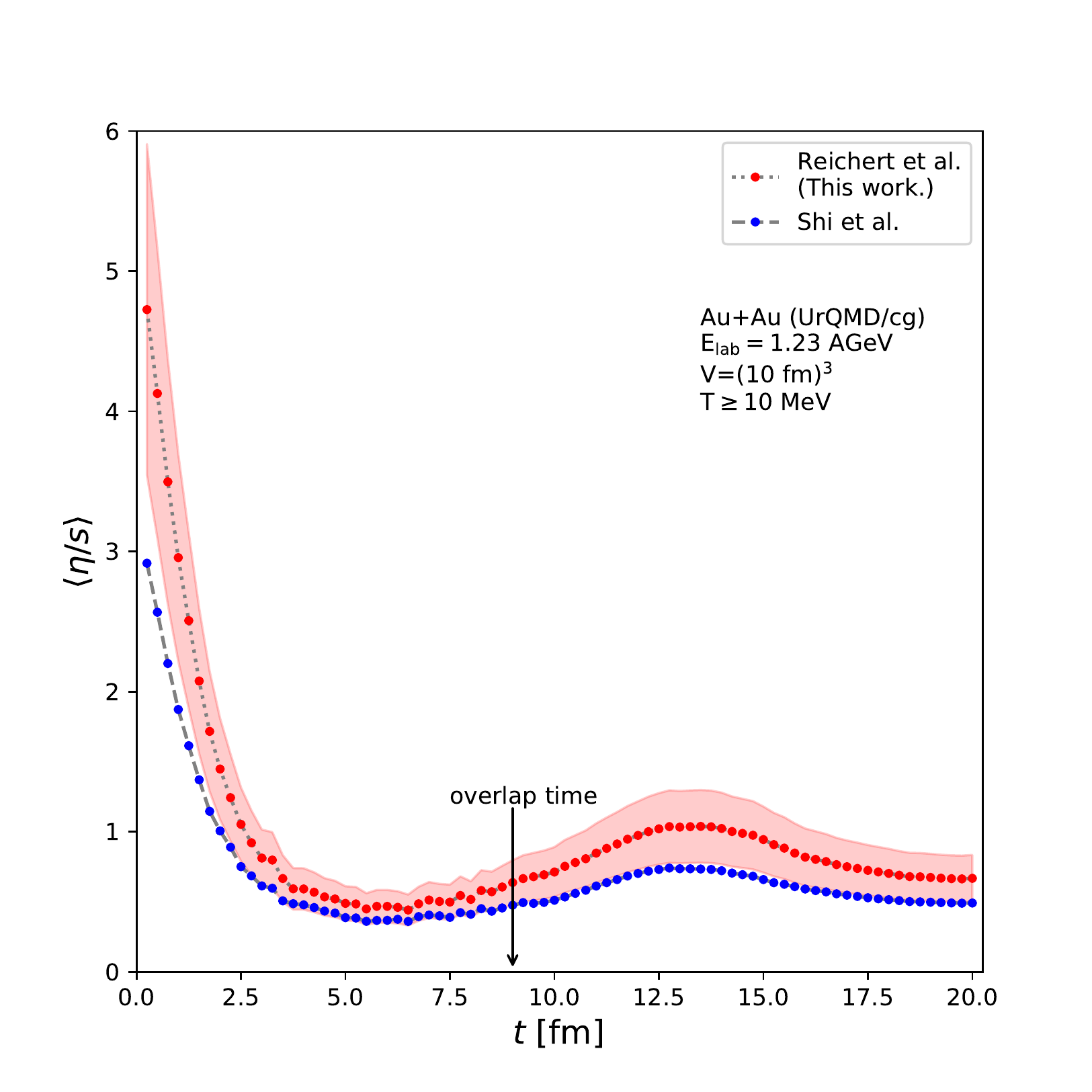}
	\caption{[Color online] Time evolution of $\langle\eta/s\rangle$ in central Au+Au reactions at a beam energy of 1.23 $A$GeV.}
	\label{etastime}
\end{figure}

Let us start by exploring the time evolution of the average% 
\footnote{We define $\langle\eta/s\rangle$ as the volume average of $\eta/s(t,\mathbf{r})$ at each time: $\langle\eta/s\rangle(t)=V^{-1}\int_{V}{\rm d}^3r\dfrac{\eta}{s}(t,\mathbf{r})$ with $V=(10\,\mathrm{fm})^3$.}
$\langle\eta/s\rangle$ ratio in the central cube, $V=(10\,\mathrm{fm})^3$, for Au+Au reactions at $E_{\rm lab}=1.23$~$A$GeV as shown in Figure \ref{etastime}. The red line shows our calculation using Eq. \ref{eta_kinetic} and with the parameters discussed above  \cite{Danielewicz:1984kt,Torres-Rincon:2012sda}, the red error band is determined by the difference of $s$- and $p$-wave scatterings. The blue line denotes the results employing the parametrization by Shi and Danielewicz given in \cite{Shi:2003np} assuming full equilibrium. This comparison further supports that the assumption of local equilibrium is fulfilled to a very good degree. 

We further observe that $\eta/s$ has a pronounced time dependence, which reflects the temperature and density evolution of the system. I.e. during the initial stage of the nuclear medium the shear viscosity to entropy density ratio is large, but rapidly decreasing, reflecting the increase in temperature and baryon density. After the initial heating, $\eta/s$ develops a plateau in line with a rather constant temperature and baryon chemical potential (in line with \cite{Steinheimer:2016vzu}) until full overlap ($t=9$~fm) where expansion takes over and $\eta/s$ increases again slightly. This weak time dependence of $\eta/s$ after the initial collision allows us to extract a value for the viscosity ratio at this energy.
 
\begin{figure} [t!hb]
	\includegraphics[width=\columnwidth]{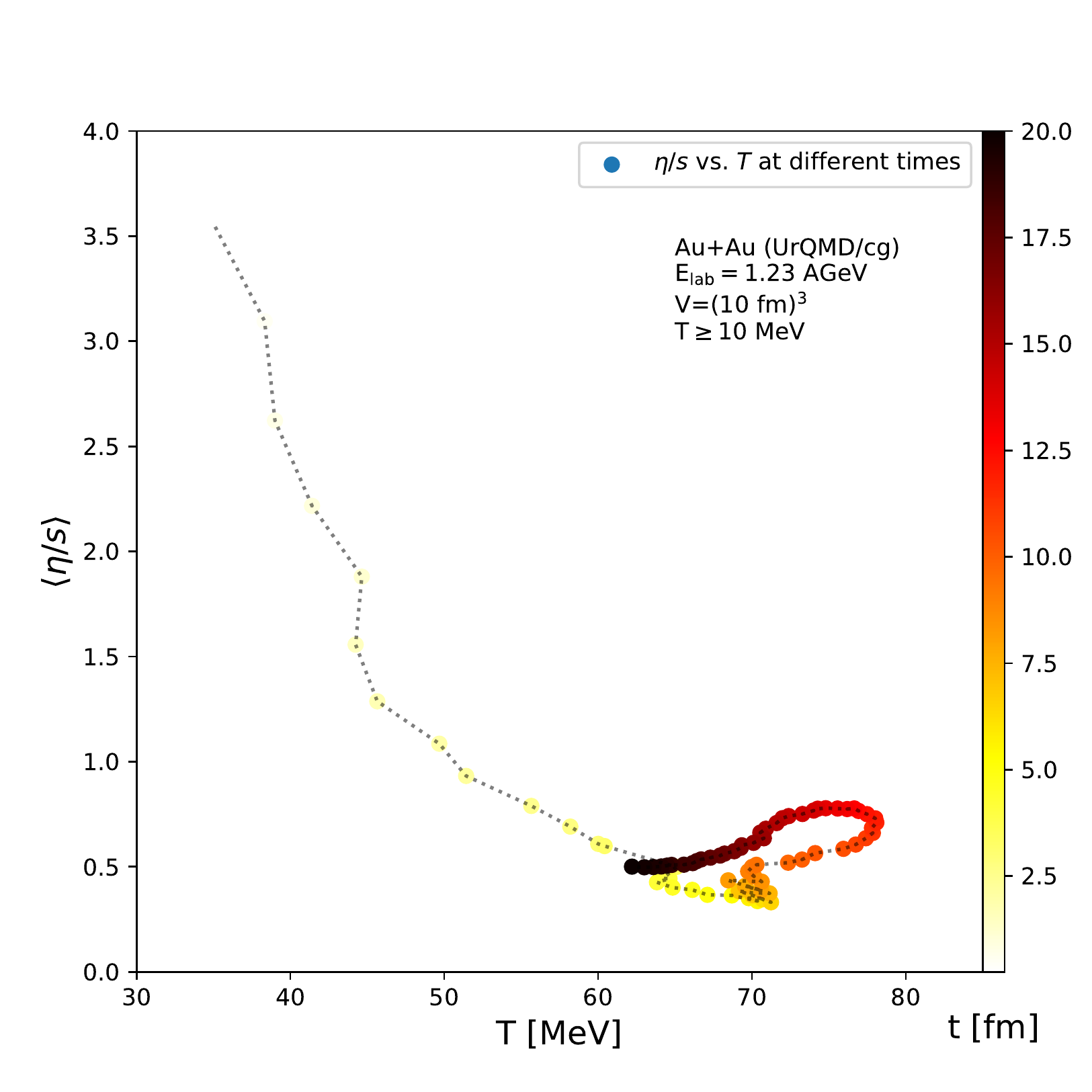}
	\caption{[Color online] Evolution of $\langle\eta/s\rangle$ in central Au+Au reactions at a beam energy of 1.23 $A$GeV as a function of temperature $T$. The time evolution is encoded in the color, see scale on the right hand side.}
	\label{etastemp}
\end{figure}

Next we turn to the temperature evolution. Figure \ref{etastemp} shows $\langle\eta/s\rangle$ in dependence of the temperature, again averaged over the central cube with side lengths of 10~fm. The time at which the average is calculated is denoted by the color scale on the r.h.s. 
We observe a sharp decrease of $\langle\eta/s\rangle$ during the compression stage in line with rising temperature. In this phase, the entropy production increases stronger than the shear viscosity. During the relevant overlap time, the temperature stays approximately constant at $60-80$~MeV which corresponds to the time of maximal compression as already seen in the time evolution in Fig. \ref{etastime}. Finally, after $\approx 15$~fm the expansion rate becomes the driving force and hence the system cools down.

\begin{figure} [t!hb]
	\includegraphics[width=\columnwidth]{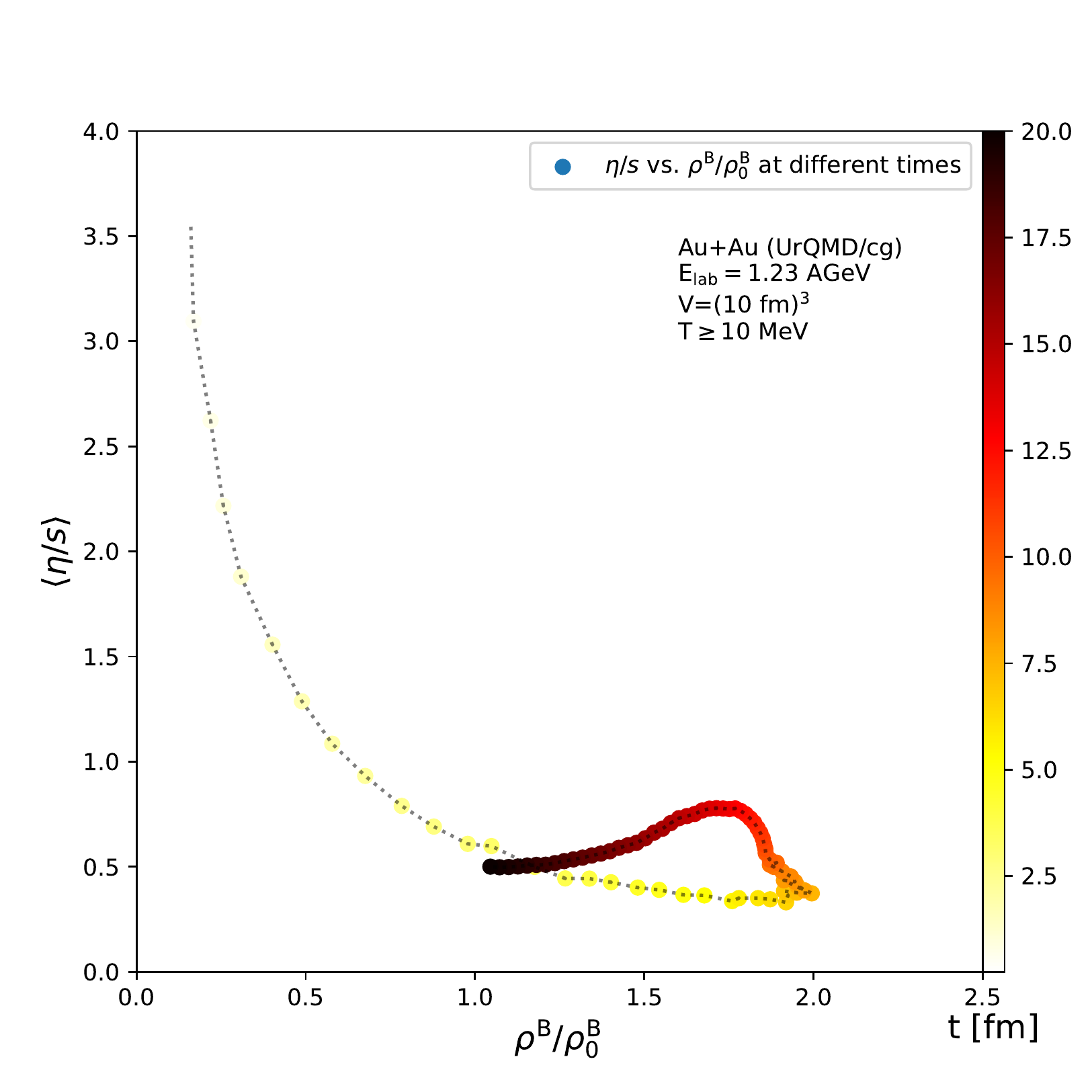}
	\caption{[Color online] Evolution of $\langle\eta/s\rangle$ in central Au+Au reactions at a beam energy of 1.23 $A$GeV as a function of the scaled baryon density $\rho_B/\rho_0$. The time evolution is encoded in the color, see scale on the right hand side.}
	\label{etasrho}
\end{figure}

Figure \ref{etasrho} shows the evolution of $\langle\eta/s\rangle$ in the same volume as a function of baryon density (normalized to the ground state density). Here one observes again an intricate time evolution: The viscosities in the compression stage and the expansion stage behave differently due to different $T(\varepsilon,\rho_{\mathrm{B}})$ distributions. During the most dense phase (from 5-10~fm) with $\rho_\mathrm{B}/\rho_0\approx2$ the system stays in the region with approximately constant $\eta/s\approx0.4-0.6$ (or $(5-7)\,(4\pi)^{-1}$). This is the stage at which the first order flow component $v_1$ is mostly generated. As the HADES data \cite{Adamczewski-Musch:2020iio} revealed, the slope of $\mathrm{d}v_1/\mathrm{d}y$ at midrapidity is positive at 1.23~$A$GeV beam energy. It is worth mentioning that the bulk viscosity might have a non negligible influence on $v_1$ as well \cite{Rose:2020lfc}. The second order flow component $v_2$, in contrast, is determined by an intricate interplay of the compression phase, the blocking of spectator nucleons and the subsequent onset of expansion. In turn that suggests that the $v_1$ component of the flow which is mostly generated during the compression and overlap stage receives a different contribution from the viscous corrections than the $v_2$ component which also achieves a major contribution from the late expansion stage of the system. As pointed out in Ref. \cite{Adamczewski-Musch:2020iio}, further insights into the interplay of shear (and bulk) viscosity and the Fourier-decomposition of azimuthal particle distribution may be investigated through higher order flow components and their correlations.

\subsection{$\eta/s$ in the transverse plane}
\begin{figure} [t!hb]
	\includegraphics[width=\columnwidth]{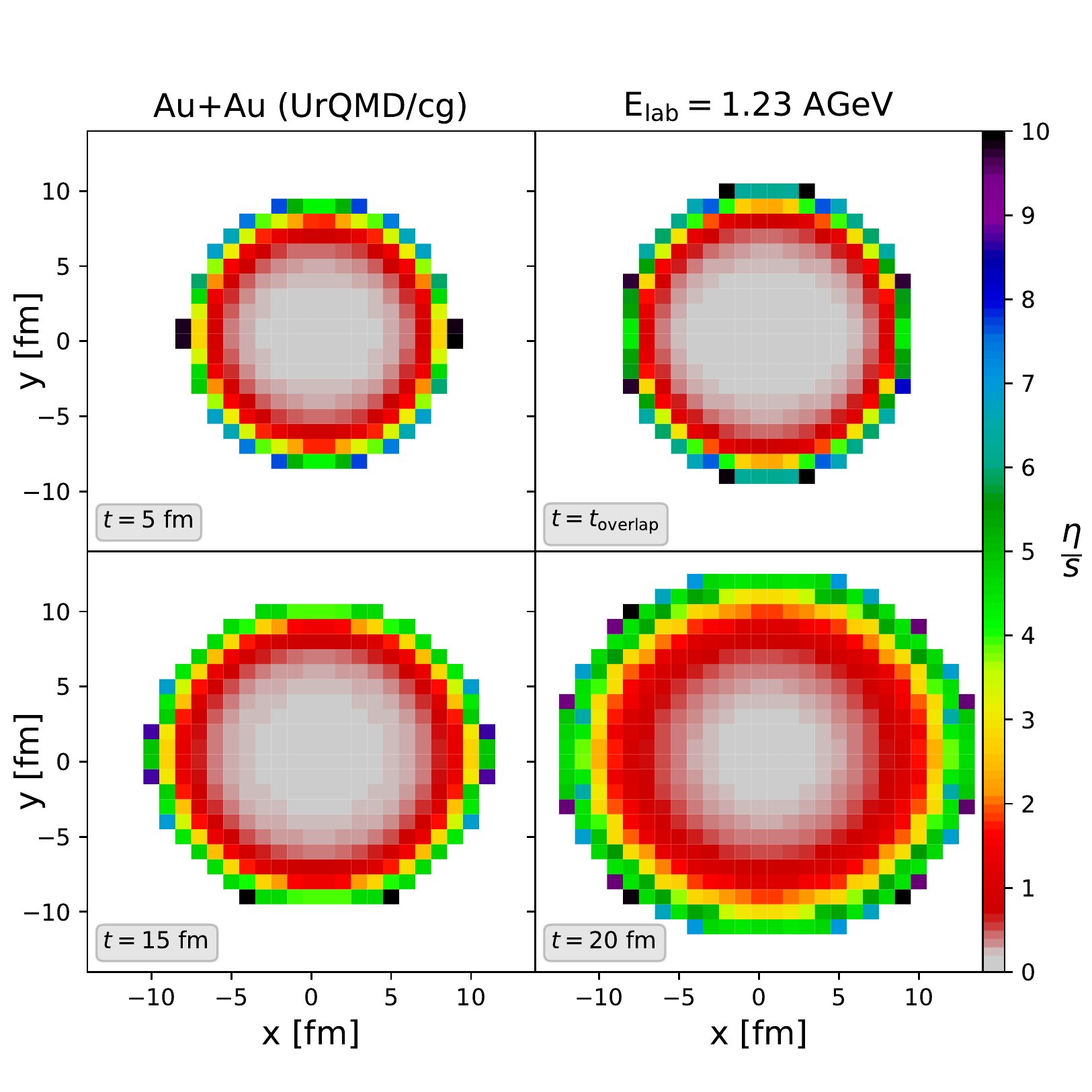}
	\caption{[Color online] Distribution of $\eta/s$ in cells in the transverse (x-y) plane with $-1\leq z\leq1$~fm at t=5~fm (top left), $t_\mathrm{overlap}$ (top right), 15~fm (bottom left) and 20~fm (bottom right) in central Au+Au reactions at a beam energy of E$_\mathrm{lab}=1.23$~$A$GeV from UrQMD. The $\eta/s$ values are encoded in the color, see scale on the right hand side.}\label{etaS_transverse_z0}
\end{figure}

Now that the difference between the compression phase and the expansion phase in central Au+Au reactions at 1.23 $A$GeV beam energy is elaborated, we will take a look into $\eta/s$ in the transverse plane%
\footnote{The average is taken over the interval $z\in[-1,\,1]$ for every x-y cell at the given times.} 
at $z\approx 0$ and at different times of the reaction. Fig. \ref{etaS_transverse_z0} shows the distribution of $\langle\eta/s(x,y)\rangle$ in the transverse plane (x-y plane) for $z=0$~fm. The top left figure shows the beginning of the reaction ($t=5$~fm), the top right figure indicates the $\eta/s$ distribution at the time of maximal compression (i.e. at $t \approx10$~fm), while the bottom figures show the transverse distribution of $\eta/s$ in the expansion stage, i.e. after 15~fm and 20~fm. Here we observe that the shear viscosity is rather uniformly distributed within the transverse plane. However, with further distance from the center of the fireball the $\eta/s$ ratio increases rapidly. As expected, the region of low $\eta/s$ evolves with elapsing time from an elliptic to a circular shape. As a side remark, we want to point out that the spatial gradients in $\eta/s$ might influence the evolution of the vorticity $\bm{\omega}$ via the vorticity transport equation ${\rm D}\bm{\omega}/{\rm D}t=(\bm{\omega}\cdot{\bm\nabla})\mathbf{u}+\eta{\bm \nabla}^2\bm{\omega}$, with $\mathbf{u}$ being the fluid velocity. This may allow to use $\Lambda$ and $\bar{\Lambda}$ polarization measurements to explore the radial dependence of $\eta/s$ using the different spatial emission regions of $\Lambda$ and $\bar{\Lambda}$ as suggested in \cite{Ayala:2020soy}. 

\subsection{Energy dependence}
Finally, we put the present result at very low energies in perspective to the existing estimates of $\eta/s$ at higher collision energies. 

\begin{figure} [t!hb]
	\includegraphics[width=\columnwidth]{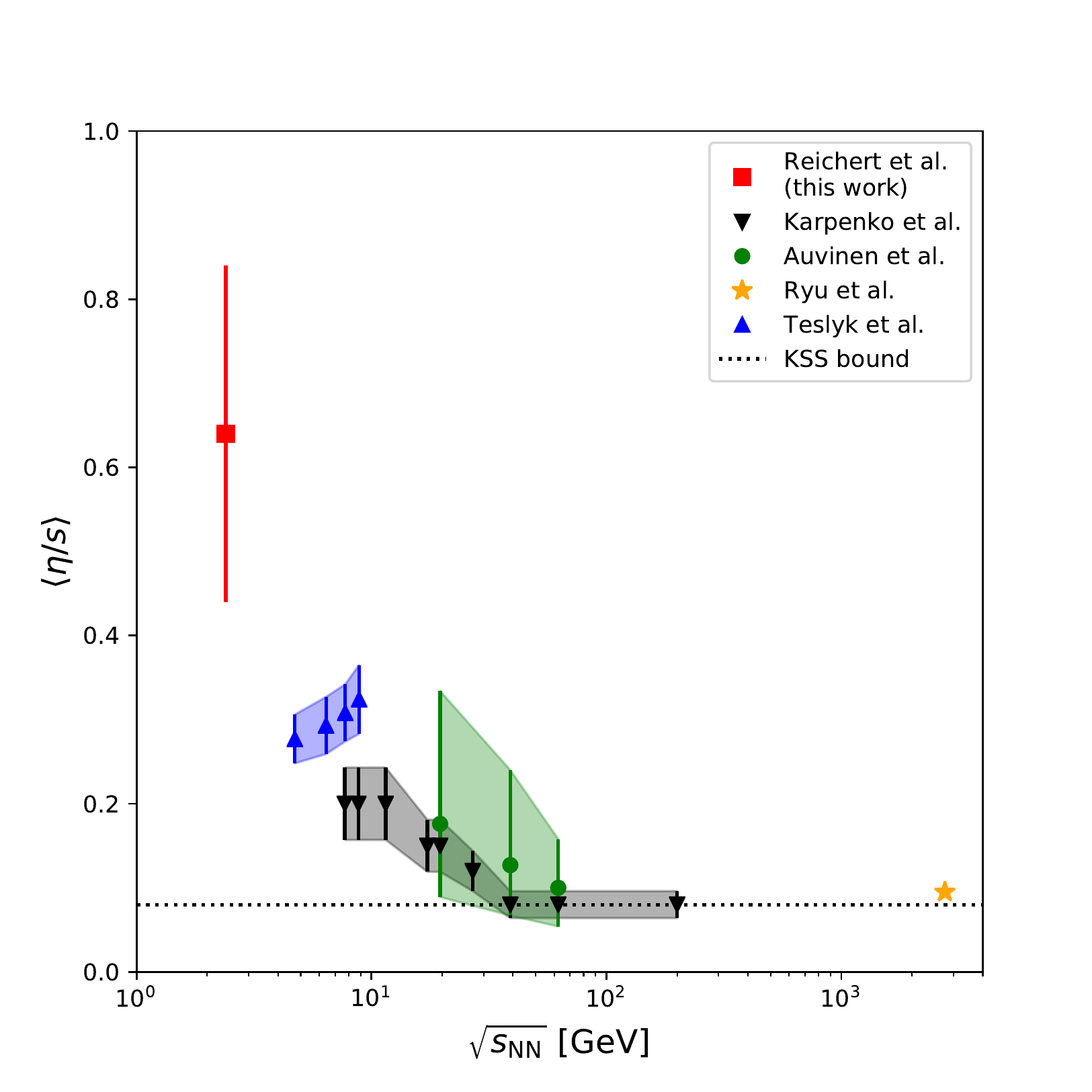}
	\caption{[Color online] Dependence of $\eta/s$ from $\sqrt{s_\mathrm{NN}}$. The red square (this work) is taken at full overlap. The blue triangles-up are extracted from \cite{Teslyk:2019ioo} at full overlap, while the black triangles-down are from \cite{Karpenko:2015xea} and the green circles are from \cite{Auvinen:2017fjw}. The orange star denotes an estimate from \cite{Ryu:2015vwa}, and the dotted black line shows the KSS boundary \cite{Policastro:2001yc,Kovtun:2004de}.}\label{etaS_sNN}
\end{figure}

In Fig. \ref{etaS_sNN} we show the dependence of $\eta/s$ on the center of mass energy $\sqrt{s_\mathrm{NN}}$ in nucleus-nucleus collisions. The red square (this work) is calculated at the time of full nuclear overlap. The error bar stems from the upper and lower limits estimated from the $s$- and $p$-wave scaling. The blue triangles-up are extracted from \cite{Teslyk:2019ioo} at full overlap, while the black triangles-down are from \cite{Karpenko:2015xea} and the green circles are from \cite{Auvinen:2017fjw} both employing a Bayesian analysis of elliptic flow data from RHIC in comparison to hydrodynamic simulations. The orange star denotes an estimate from hydrodynamic calculations at LHC \cite{Ryu:2015vwa}, while the dotted black line shows the KSS boundary \cite{Policastro:2001yc,Kovtun:2004de}.
We observe that a) the extracted $\eta/s$ values at high energies (including error bars) approach the KSS bound for strongly coupled systems at $\sqrt{s_{\rm NN}}$ around 40~GeV and that b) in the GSI/FAIR energy regime the $\eta/s$ values increase strongly with decreasing center-of-mass energy reaching typical values for a rather cold relativistic hadron gas. 

\section{Summary}
In this article we have used the Ultra-relativistic Quantum Molecular Dynamics (UrQMD) transport approach combined with a coarse graining procedure to extract for the first time the shear viscosity to entropy density ratio in Au+Au reactions in the GSI/FAIR energy regime. The extracted value in Au+Au reactions at $\sqrt{s_\mathrm{NN}}=2.4$~GeV is $\langle \eta/s\rangle=0.65\pm0.15$. We showed that the $\eta/s$ ratio is space and time dependent. We found that $\eta/s$ is decreasing rapidly during the compression phase and levels of around the maximum compression of $\rho_B/\rho_0=$2, corresponding to a temperature of $T=60-80$~MeV. We further showed the distribution of $\eta/s$ in the transverse plane and discussed how the spatial distribution of $\eta/s$ might be explored with $\Lambda$, $\bar{\Lambda}$ polarization measurements. We suggest that at the discussed FAIR energies different flow components receive different contributions from the shear viscosity to entropy density ratio because they are created at different times of the collision. I.e. $v_1$ is mainly created during the maximal compression phase where $\eta/s$ is lowest, while $v_2$ gains further contributions during the expansion phase where $\eta/s$ moderately increases. Finally, we related our result to previous estimates on $\eta/s$ as a function of $\sqrt{s_{\rm NN}}$ showing that the extracted value of $\eta/s\approx0.65$ carries on the trend established by previous calculations at higher energies.

\begin{acknowledgements}
The authors thank Behruz Kardan from the HADES collaboration for stimulating discussion at the "HIPSTARS - Workshop on Heavy Ion Physics and Compact Stars" workshop held online in December 2020. This work was supported by Helmholtz Forschungsakademie Hessen (HFHF), and in the framework of COST Action CA15213 THOR. G. Inghirami is supported by the Academy of Finland, Project no. 297058. Computational resources were provided by the Center for Scientific Computing (CSC) of the Goethe University. 
\end{acknowledgements}

%%%%%%%%%%%%%%%%%%%%%%%%%%%%%%%%%%%%%%%%%%%%%%%%%%%%%%%%%%%%%%%%%%%%%%%%%%%%%%%

%%%%%%%%%%%%%%%%%%%%%%%%%%%%%%%%%%%%%%%%%%%%%%%%%%%%%%%%%%%%%%%%%%%%%%%%%%%%%%% 
\end{document}